\documentclass[prb,twocolumn,showpacs,preprintnumbers,amsmath,amssymb]{revtex4}
\usepackage{graphics}
\usepackage{graphicx}
\usepackage{color}

\newcommand{\Ns}{\ensuremath{N_{\mathrm{SB}}}}
\newcommand{\EF}{\ensuremath{E_F}}
\newcommand{\vc}{\vec}
\newcommand{\kpar}{\ensuremath{\vc{k}}_{\parallel}}

\begin{document}

\title{Spin injection from Fe into Si(001): {\it ab initio}
  calculations and role of the Si complex band structure}

\author{Phivos Mavropoulos}

\affiliation{Institut f\"ur Festk\"orperforschung, Forschungszentrum
  J\"ulich, D-52425 J\"ulich, Germany}

\begin{abstract}
We study the possibility of spin injection from Fe into Si(001), using
the Schottky barrier at the Fe/Si contact as tunneling barrier. Our
calculations are based on density-functional theory for the
description of the electronic structure and on a Landauer-B\"uttiker
approach for the current. The current-carrying states correspond to
the six conduction band minima of Si, which, when projected on the
(001) surface Brillouin zone (SBZ), form five conductance hot spots:
one at the SBZ center and four symmetric satellites. The satellites
yield a current polarization of about 50\%, while the SBZ center can,
under very low gate voltage, yield up to almost 100\%, showing a
zero-gate anomaly. This extremely high polarization is traced back to
the symmetry mismatch of the minority-spin Fe wavefunctions to the
conduction band wavefunctions of Si at the SBZ center. The tunneling
current is determined by the complex band structure of Si in the [001]
direction, which shows qualitative differences compared to that of
direct-gap semiconductors. Depending on the Fermi level position and
Schottky barrier thickness, the complex band structure can cause the
contribution of the satellites to be orders of magnitude higher or
lower than the central contribution. Thus, by appropriate tuning of
the interface properties, there is a possibility to cut off the
satellite contribution and to reach high injection efficiency. Also,
we find that a moderate strain of 0.5\% along the [001] direction is
sufficient to lift the degeneracy of the pockets so that only states
at the zone center can carry current.
\end{abstract}


\maketitle

\section{Introduction}

The electrical injection of spin polarized carriers, in short called
\emph{electrical spin injection}, into the conduction band of
semiconductors, is one of the key elements for realizing spin
transistors such as the one proposed by Datta and Das.\cite{Datta90}
Following arguments by Schmidt and co-workers,\cite{Schmidt00}
Rashba,\cite{Rashba01} and Fert and Jaffr\`es,\cite{Fert01} it was
realized that efficient spin injection requires the presence of a
tunnel barrier at the ferromagnet/semiconductor interface. Significant
experimental success started in 2001-2002,\cite{Hanbicki02} in
junctions that contained a tunnel barrier between the ferromagnetic
metal and the semiconductor, either in the form of the Schottky
barrier of the semiconductor itself or in the form of an ultra-thin
film of some other insulating material,\cite{Erve04} such as AlO or
MgO.\cite{Jiang05} Most works have focussed on injection into
GaAs,\cite{Hanbicki02,Erve04,Jiang05} where the current polarization
can be optically detected in GaAs/AlGaAs/GaAs quantum wells. Recently,
however, electrical spin injection into Si has also been demonstrated
both via electrical spin detection\cite{Appelbaum07,Huang07a,Huang07b}
and optical detection\cite{Jonker07} with impressively large spin
coherence length, reaching up to 350~$\mu$m.\cite{Huang07b}

The present work was motivated by the increasing experimental activity
in the direction of spin injection and manipulation in
Si,\cite{Appelbaum07,Huang07a,Huang07b,Jonker07,Min06,Uhrmann08} as well as by
arguments for advantages of spin transport in Si.\cite{Zutic06} Based
on density-functional calculations we examine the possibility of
direct spin injection from Fe into Si, with the Schottky barrier of Si
used as the necessary tunneling barrier. We focus on idealized
epitaxial, atomically flat Fe/Si(001) interfaces. After briefly
presenting the magnetic structure of the Fe/Si interface, we discuss
the tunneling properties of Si based on the concept of complex band
structure. We then show that spin injection is possible and can be
tuned by adjusting the Fermi level \EF\ in the band gap of Si in the
tunneling region, or by straining Si in the injection region in order
to cut off satellite contributions. We further discuss the so-called
$\bar{\Gamma}$-point rule for increased injection efficiency and show
that, in the case of Fe/Si, it is harder to satisfy than in
Fe/GaAs.

\section{Method and details of calculation. Limitations of approach \label{sec:method}}

The electronic structure of the junctions was calculated within the
local (spin) density approximation, L(S)DA, to density functional
theory. The Kohn-Sham equations were solved using the screened
Korringa-Kohn-Rostoker (KKR) Green function method.\cite{tbkkr} A
non-relativistic treatment was chosen, as it is known that
relativistic corrections result in a reduction of the calculated
semiconductor gaps within the LDA. The atomic sphere approximation was
used in most cases, except in particular tests which were performed
with a full potential treatment. The conductance was calculated in the
zero-bias limit within a Landauer-B\"uttiker approach adjusted to the
KKR Green function method.\cite{Mavropoulos04}

The self-consistent electronic structure of the junction was
calculated using the decimation technique\cite{decimation} for the consideration of the
half-infinite leads. An angular momentum cutoff of
$\ell_{\mathrm{max}}=2$ was used. The details of the electronic
structure that are important here change only little between
$\ell_{\mathrm{max}}=2$ and $\ell_{\mathrm{max}}=3$; for example, the
spin moment of bcc Fe changes by less than 5\%. However, an angular
momentum cutoff of $\ell_{\mathrm{max}}=3$ was
necessary\cite{Mavropoulos04} for the calculation of the conductance
matrix elements.

The Fe/Si(001) interface was taken to be ideal (atomically
flat). Contrary to the wavefunctions, the perturbation of the
potentials dies out relatively fast with the distance from the
interface. Within a finite region of two atomic layers around the
interface, the potentials are perturbed from their bulk-like values and
calculated self-consistently by sandwiching 5 layers of Si between two
Fe leads. This way also provides the band alignment between Fe and
Si. Subsequently the half-infinite Si lead is constructed by
maintaining the potentials only up to the middle of the junction (left
Fe lead and Si up to third layer), and repeating the potential of the
third Si layer (which is considered already bulk-like) ad infinitum by
use of the decimation technique. The decimation technique is also used
to produce the half-infinite, bulk-like Fe lead. The same method has
been applied in previous works on Fe/GaAs and
Fe/ZnSe.\cite{Wunnicke02,Wunnicke04}. 

\begin{figure}
\begin{center}
\includegraphics*[width=8cm]{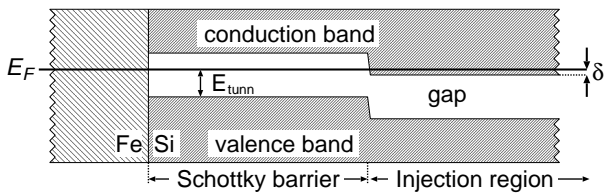}
\caption{Schematic representation of the band diagram at the Fe/Si
  interface used in the calculations.\label{fig:banddiagram}}
\end{center}
\end{figure}

We consider the following external parameters that are involved in an
injection experiment (see also Fig.~\ref{fig:banddiagram}). Firstly,
the Schottky barrier has a thickness of \Ns\ layers and a height
determined by the band alignment of Si with respect to the Fermi level
of Fe. Experimentally, this can be adjusted by appropriate doping or
by a gate voltage.  In the calculations, this is simulated by adding
an appropriate constant shift to the Si potentials up to \Ns, so that,
within the barrier, the bands are shifted and \EF\ (which is fixed by
the Fe lead) falls at a desired position between the valence band and
conduction band edges, $E_v$ and $E_c$: $\EF=E_v+E_{tunn}$. In this
step the first two Si layers at the interface are excluded, as it is
considered that their electronic structure will be mainly affected by
their proximity to Fe. Secondly, after the barrier region the band
alignment must be such that the injected electrons arrive in the Si
valence band. Experimentally the appropriate band alignment can also
be adjusted by a gate voltage or by doping. Here, it is again
accounted for by adding a proper energy constant to the potentials of
all Si layers from $\Ns+1$ and on, so that \EF\ falls slightly in
above $E_c$: $\EF=E_c+\delta$. In the present work $\delta=28$~meV
(2~mRyd) was used except where otherwise indicated.

A few comments are in order on the chosen band diagram. Firstly, in
experiment a Schottky barrier will possibly have a different, less
abrupt shape. We defer this question for Sec.~\ref{sec:summary},
because after the discussion has revealed the effect of the
positioning of \EF\ in the gap. Secondly, the chosen value of
$\delta=28$~meV can be perhaps reasonable for GaAs, but would probably
result in an unrealistically high carrier concentration in Si due to
the high Si effective mass. This choice helps in visualizing and
understanding the current distribution in the SBZ. In
Sec.~\ref{sec:conductance} we also show calculations with $\delta$ up to
1.4~meV, which in fact reveal a qualitative change of injection
efficiency.

It is well known that local density-functional theory (as is the LDA
or the generalized gradient approximation) underestimates the gap in
semiconductors and insulators; in the present calculations the Si band
gap is found to be approximately 0.38~eV. Thus, the values of
tunneling conductance obtained within density-functional calculations
can only provide qualitative understanding, including trends, but not
quantitatively correct results. Qualitatively, the underestimated band
gap in the tunneling region can be partly compensated by increasing
the barrier thickness \Ns. Here, \Ns\ was varied up to 60 Si layers.

Especially for the calculation of the bulk Si band structure
(including the complex band structure), significant quantitative
improvement can be achieved by using an orthogonalized plane wave
method with pseudopotentials fitted to optical excitation
experiments. Here we followed such a method with the parameters taken
from Ref.~[\onlinecite{Cohen66}]; in the presentation of the results we
indicate when this method has been used.

Furthermore we note that an atomically flat interface is, at this
point, an idealization. In an experiment, most likely interface
roughness and some form of iron silicide will be present at the
interface. However, it cannot be excluded that good quality,
atomically almost flat interfaces can, in principle, be made, as has
been achieved in Fe/MgO junctions. Moreover, the parameter space
becomes simply too big if many possibilities of interface structure
are to be calculated. Since modern experimental techniques allow for a
detailed imaging of the interface structure,\cite{Zega06} it is
possible to relate further calculations to such experimental input. In
Sec.~\ref{sec:summary} we point out which calculated properties are
specific to the idealization of a flat interface.

Finally we comment on the lattice mismatch of Fe and Si. The Si
lattice constant (5.43 \AA) is approximately twice the one of Fe (2.87
\AA) with a mismatch of about 5\%. We adopt the Si lattice constant
for the calculation (the in-plane unit cell accommodates now two Fe
atoms per Fe layer), to simulate the situation where a thin film of Fe
is in contact with a thick Si barrier.  The stress on the Fe contacts
will be appreciable, but it is sufficient that Fe retains the
prescribed epitaxial structure for only a few layers for the our
conclusions to be correct. This has been demonstrated in calculations
on Fe/MgO/Fe junctions,\cite{Heiliger07} where it was shown that the
transport properties depend on the structure of the Fe lead only close
to the interface: even if Fe becomes amorphous a few layers away from
the interface, the spin filtering properties practically do not
change. However, the choice of lattice constant does affect the
moments significantly, and therefore tests on interface relaxation and
tetragonalization were made, as we discuss in Sec.~\ref{Sec:Moments}.

\section{Spin moments at the interface. Effect of lattice
  constant \label{Sec:Moments}}

\begin{figure}
\begin{center}
\includegraphics[width=8cm]{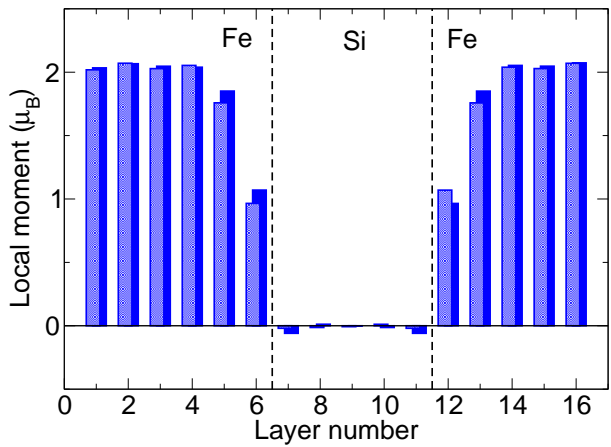}
\caption{(color online) Atom-resolved spin moments close to the
  interface in a Fe/Si/Fe(001) junction, calculated at the Si lattice
  parameter. Evidently the Fe moments at the interface are strongly
  suppressed. The two values for each layer correspond to the two
  inequivalent atoms.\label{fig:moments}}
\end{center}
\end{figure}

The spin moment of bulk Fe was found not to change too much with
lattice constant: a moment reduction of 12\% was calculated at the Si
lattice constant compared to the result at the Fe lattice
constant. However, the change of the moments at the interface is
appreciable. At the Si lattice constant, the moments of the interface
Fe atoms are reduced to approximately half their bulk value, as is
shown in Fig.~\ref{fig:moments}. This effect, previously reported,
e.g., in Ref.~[\onlinecite{Freyss02}], is not present in calculated
junctions at the Fe lattice constant, where the interface Fe moments
are not suppressed at all.

The magnetic structure was further tested by calculating a finite slab
of 4 layers Fe/ 4 layers Si / 4 layers Fe with a full potential
treatment and $\ell_{\mathrm{max}}=3$. Here the in-plane lattice
constant of Si was adopted, while the Fe layers were allowed to assume
an optimal $c/a$ ratio, and the interlayer distance at the interface
was relaxed. No qualitative differences were found compared to the
results using the Si lattice constant for $a$ and $c$: the strong
suppression of the moments was still present.

We conclude that a strong suppression of the Fe interface moment is
unavoidable if the in-plane lattice parameter of Si(001) is
adopted. Interestingly, this does not qualitatively affect the
transport properties that interest us here, in particular the spin
polarization of the injected current. This was found by test
calculations of spin injection in a Fe/Si(001) junction using the Fe
lattice parameter (whence the moments at the interface where not
compromised).

\section{Role of Silicon real and complex band structure
  \label{Sec:bandstr}}

Silicon has an indirect band gap. The valence band has a maximum at
the $\Gamma$ point ($\vc{k}=0$), while the conduction band has six
equivalent degenerate minima at $\vc{k}=(\pm k_0,0,0),\ (0,\pm
k_0,0),\ (0,0,\pm k_0)$ with $k_0\approx 0.85\ 2\pi/a$ ($a$ is the
lattice parameter). These properties are most important for a
qualitative discussion of the tunneling conductance and electron
injection. 

\begin{figure}
\begin{center}
\includegraphics[width=8.5cm]{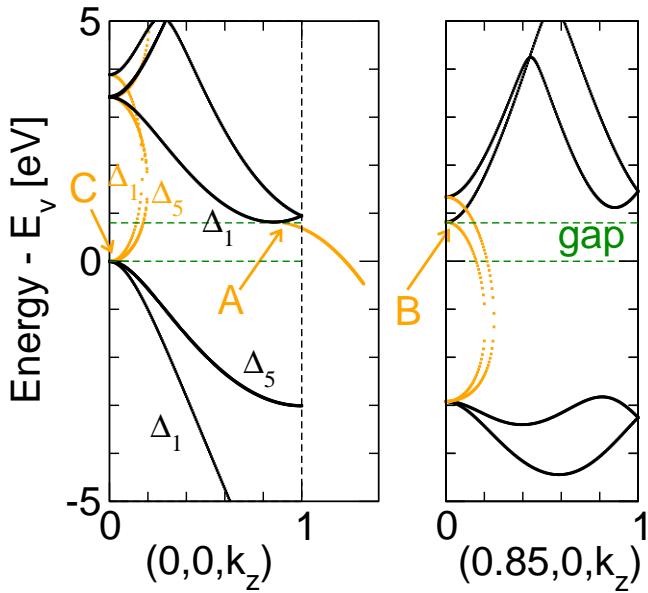}
\caption{(color online) Real and complex band structure of Si along
  the $k_z$ direction at $\kpar=0$ (left) and $\kpar=(0.85,0)\
  (2\pi/a)$ (right). Black lines stand for real bands as $E=E(k_z)$;
  orange (or grey) lines stand for complex bands as $E=E(k_z + i
  \kappa)$. The effective mass anisotropy at the six pockets of the
  conduction band edge, $k_{x,y,z}\approx\pm 0.85\ (2\pi/a)$ evidently
  results in a quite different curvature of the complex bands
  departing from these (indicated by arrows). Therefore, when \EF\ is
  close to the conduction band edge, the tunneling is dominated by the
  contributions at the four equivalent points $\kpar\approx(\pm
  0.85,0)\ (2\pi/a)$, $\kpar\approx(0,\pm 0.85)\ (2\pi/a)$. The
  symmetry of the bands is also indicated ($\Delta_1$ and
  $\Delta_5$). (More complex bands, irrelevant for the discussion,
  have been omitted.)
  \label{fig:bandstr}}
\end{center}
\end{figure}

\begin{figure}
\begin{center}
\includegraphics*[width=8cm]{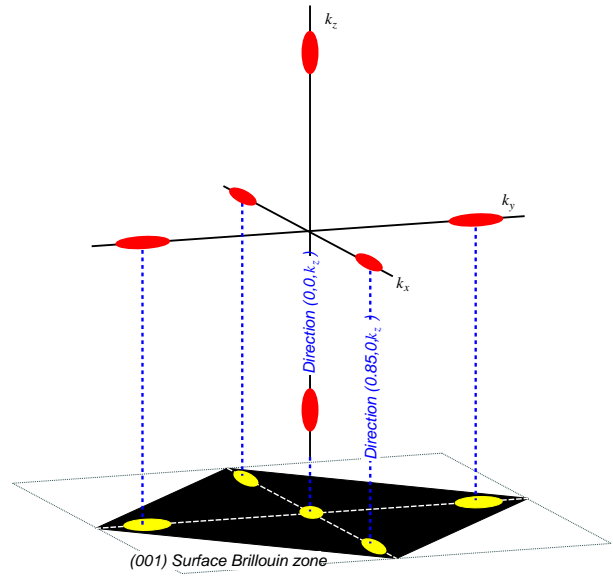}
\caption{(color online) Schematic representation of the six equivalent
  Si conduction band ``pockets''. When these are projected on the
  two-dimensional (001) surface Brillouin zone, five hot
  spots for conductance in the [001] direction arise: one central (at
  $\kpar=0$) and four equivalent satellites. \label{fig:Si_pockets}}
\end{center}
\end{figure}

The results of this section were obtained by using pseudopotentials
fitted to optical transitions.\cite{Cohen66} The band structure of Si
around the gap is shown in Fig.~\ref{fig:bandstr}. The conduction band
around the minima forms six ellipsoidal ``pockets'' of highly
anisotropic effective mass, as shown schematically in
Fig.~\ref{fig:Si_pockets}. Electron injection takes place into these
pockets; if the junction growth is along the [001] direction, then the
pockets are projected in five conductance hot spots on the
two-dimensional (001) surface Brillouin zone
(Fig.~\ref{fig:Si_pockets}). One of these is at $\kpar=0$, while the
other four are equivalent and form a satellite structure. (We denote
by $\kpar$ the projection of the Bloch $\vc{k}$-vector on the surface
Brillouin zone.)

The pocket structure is highly important also for the tunneling
properties. Setting the $z$ axis along the [001] epitaxy direction,
the tunneling wavefunction at $\EF$ depends on the decay parameter
$\kappa(\kpar;\EF)$ as $\psi(\vc{r})\sim\exp(-\kappa \,z)$. The set of
$\kappa(\kpar;E)$ forms the so-called complex band structure;
evidently, the lowest values of $\kappa$ are responsible for the
highest tunneling current. The value of $\kappa(\kpar;\EF)$ depends on
the proximity of \EF\ to the Bloch eigenvalues $E(\kpar;k_z)$ \emph{of
the same $\kpar$} in the conduction or valence band, and on the
curvature of $E(\kpar;k_z)$ with respect to $k_z$ (i.e., the effective
mass in the $k_z$ direction). Thus the effective mass anisotropy of
the pockets in the Si tunneling barrier has consequences that we now
discuss.

Fig.~\ref{fig:bandstr} shows the complex (and real) band structure in
the [001] ($k_z$) direction for two different values of $\kpar$:
$\kpar=0$ and $\kpar=(k_0,0)$; the latter corresponds to one of the
aforementioned ``satellite'' conductance hot spots. Arrows A and B
indicate the pocket positions, and the effective mass anisotropy is
evident. The indicated pocket in the left panel (arrow A) has its long
axis along the $k_z$ direction. The complex band (in orange, arrow A)
departing from the edge of this pocket has a small curvature (as it is
along the long axis of the pocket), leading quickly to high values of
$\kappa$ for $\EF<E_c$. On the other hand, the pocket indicated in the
right panel (arrow B) has its long axis oriented in the $k_x$
direction. Here, the complex band along $k_z$ inherits a large
curvature from the small axis of the pocket, and the value of
$\kappa$ remains relatively small for $\EF<E_c$.

Consequently, when \EF\ is in the gap but close to the conduction band
edge, the tunneling current through the satellite positions dominates,
while the tunneling current through the Brillouin zone center is
small. However, if \EF\ is lowered closer to the valence band edge, a
different contribution from the Brillouin zone center becomes more
important, marked by arrow C in Fig.~\ref{fig:bandstr}. This comes
about via the complex band of $\Delta_1$ symmetry arising from the
valence band maximum. 

\begin{figure}
\begin{center}
\includegraphics*[width=8cm]{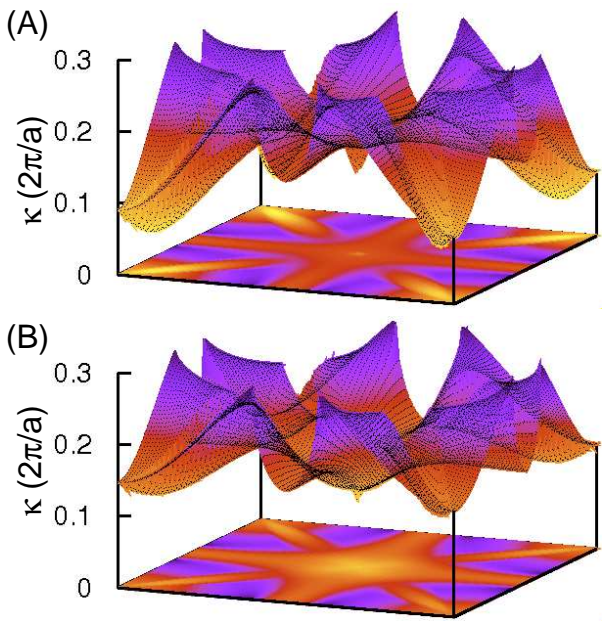}
\caption{(color online) Complex band structure (decay parameter
  $\kappa(\kpar;\EF)$) in the band gap of Si in the [001] direction
  for $\EF=E_v+0.725$~eV (A) and at $\EF=E_v+0.325$~eV (B). In
  the former case, \EF\ is close to the conduction band edge, and the
  four equivalent contributions at $\kpar\approx (\pm 0.85,0)\
  (2\pi/a)$ and $\kpar\approx (0, \pm 0.85)\ (2\pi/a)$ dominate the
  tunneling; in the latter case, \EF\ is located a little lower than
  tne middle of the gap, and the decay parameter at $\kpar=0$ is
  lowest. Only the lowest branch of $\kappa(\kpar)$ is shown. Brighter
  shaded regions correspond to lower $\kappa$ and more efficient
  tunneling.\label{fig:cplxfs}}
\end{center}
\end{figure}

The picture becomes more complete if the full (001) surface Brillouin
zone is scanned for the lowest branch of $\kappa(\kpar;\EF)$ at
different values of \EF. Fig.~\ref{fig:cplxfs} shows two such
``complex Fermi surfaces'': (A) one for $\EF=E_v+0.725$~eV (close to
$E_c$), and (B) one for $\EF=E_v+0.325$~eV (just a little lower than
the middle of the gap). Brighter colored regions correspond to lower
values of $\kappa$. In the first case, as \EF\ is close to $E_c$, the
parts of the complex Fermi surface at the satellite hot spots show the
lowest $\kappa$. As we shall see in the next section, conductance
calculations also show that these regions dominate the tunneling
current for an analogous choice of \EF. In the second case shown in
Fig.~\ref{fig:cplxfs}, the minimum of $\kappa$ is found at the
Brillouin zone center. Then the tunneling contribution at $\kpar=0$
dominates if the Schottky barrier is thick.

We close this section with the conclusion that the indirect gap of Si
lends features to the complex band structure which are qualitatively
different than direct-gap compounds, as GaAs, ZnSe, or MgO. Contrary
to all these direct-gap materials, where the complex band of
$\Delta_1$ symmetry at the Brillouin zone center gives the dominant
contribution irrespective of the exact position of \EF, in Si the
positioning of \EF\ can make a stark difference. This effect can have
consequences for all physical properties which depend on the complex
band structure. Particularly in spintronics applications it can affect
spin injection, tunneling magnetoresistance, but also ground-state
properties such as interlayer exchange coupling.\cite{Bruno94}

\section{Total and spin-dependent conductance. Current polarization
  \label{sec:conductance}} 

We now proceed to the presentation and discussion of the {\it ab
initio} results on the conductance and current polarization. We start
by commenting on the effect of the underestimation of the gap in
density functional theory. The conclusions of Sec.~\ref{Sec:bandstr}
are qualitatively still valid, but now, within the barrier, the
possible choice of $E_c-\EF$ is more limited (otherwise \EF\ will
enter the valence band). This results in an overestimation of the
relative contribution of the satellites to the current. As the
calculations show, the satellites dominate the current even when \EF\
is at the mid-gap position; hence, calculated results at mid-gap are
expected to be closer to a realistic situation of shallow tunneling
close to $E_c$. An overestimated proximity of \EF\ to the the valence
band is necessary so for the Brillouin zone center contribution to
prevail. Moreover, the decay parameters and exponential falloff of the
conductance with barrier thickness are underestimated.

\subsection{Total conductance}

\begin{figure}
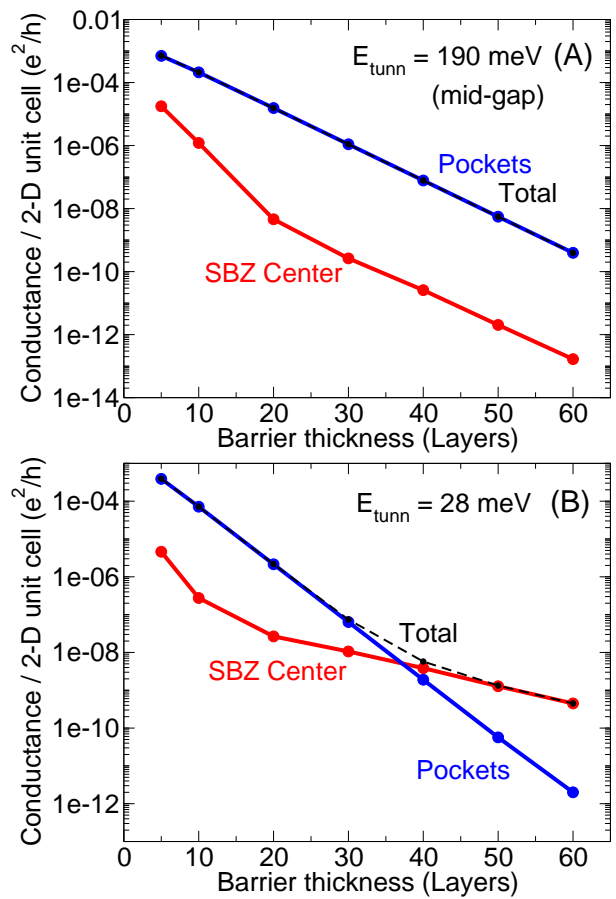

\begin{center}
\includegraphics[width=8cm]{conductance_tot_cntr_pockets_-0.022.eps}
\includegraphics[width=8cm]{conductance_tot_cntr_pockets_-0.010.eps}
\caption{(color online) Conductance per two-dimensional unit cell as a
  function of the Schottky barrier thickness for (A) mid-gap tunneling
  ($E_{tunn}=190$~meV) and (B) shallow tunneling close to the valence
  band ($E_{tunn}=28$~meV). The contributions of the region around the
  SBZ center (integrated around the $\bar{\Gamma}$-point) and the
  satellites are shown separately. \label{fig:conduct}}
\end{center}
\end{figure}

Fig.~\ref{fig:conduct} shows the calculated conductance per
two-dimensional interface unit cell as a function of the barrier
thickness. Results for two different positions of \EF\ are presented:
mid-gap tunneling ($\EF=E_v+190$~meV) and shallow tunneling close to
the valence band edge ($\EF=E_v+28$~meV). The partial contributions of
the region around the Brillouin zone center and the satellites are
also shown. An exponential decay with thickness is characteristic of
tunneling, and the slope of the curve in a logarithmic scale is
proportional to the decay parameter.

In the case of mid-gap tunneling (Fig.~\ref{fig:conduct}A) it is clear
that the satellite contribution dominates at all thicknesses, being
practically equal to the total conductance, while the central
contribution to the total current is insignificant. While the slope of
the satellite part of the conductance is constant, reflecting a single
decay parameter $\kappa_{sat}$, the central contribution gradually
changes slope; after a thickness of about 20 layers, only the
$\Delta_1$ contribution remains around $\kpar=0$, with a decay
parameter $\kappa_1\approx k_{sat}$.

The behavior in the case of shallow tunneling, closer to the valence
band edge (Fig.~\ref{fig:conduct}B), is analogous. Here, however, the
low positioning of \EF\ evidently results in $\kappa_1<\kappa_{sat}$
(see also Fig.~\ref{fig:bandstr}). Thus after a thickness of 35 layers
the central contribution prevails and dominates the tunneling current.

In both cases we also observe that, for small thicknesses (5 layers),
the satellite contribution is much stronger than the central
one. Apparently there is a much stronger coupling between the Fe and
Si states at the satellite positions in the Brillouin zone. At this
point we have no intuitive explanation for this effect.

\subsection{Spin dependent conductance and current polarization}

The spin polarization of the current is defined as
\begin{equation}
P = \frac{I_{\uparrow} - I_{\downarrow}}{I_{\uparrow} + I_{\downarrow}}
\label{eq:1}
\end{equation}
where $I_{\uparrow}$ and $I_{\downarrow}$ are the current of
majority-spin and minority-spin carriers, respectively. The difference
between $I_{\uparrow}$ and $I_{\downarrow}$ arises mainly from the
spin-dependent scattering at the Fe/Si interface (which gives rise to
spin-dependent tunneling), due to the difference in coupling of Fe
wavefunctions of different spin to Si wavefunctions at the interface. 

\begin{figure*}
\begin{center}
\includegraphics[angle=270,width=16cm]{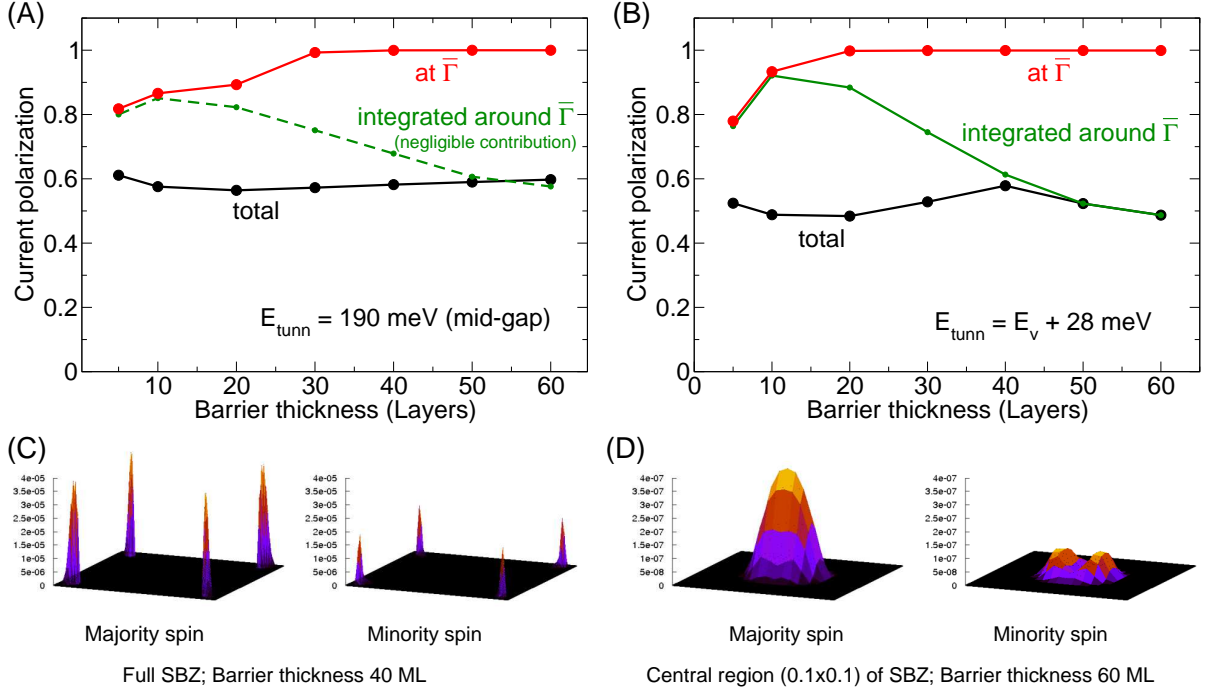}
\caption{(color online) (A and B) Current polarization as a function
  of the barrier thickness for an injection energy of $\delta=28$~meV
  and for a tunneling energy $E_{tunn}$ at mid-gap (A) and at
  28~meV (B). Shown are the total polarization (see
  Eq.~\ref{eq:1}), polarization at $\bar{\Gamma}$ (Eq.~\ref{eq:2}) and
  the integrated polarization at the central hot spot
  (Eq.~\ref{eq:3}). (C): Conductance as a function of $\kpar$ in the
  SBZ for a barrier thickness of 40 layers in the case of $E_{tunn}$
  at mid-gap. The SBZ center contribution is insignificant next to the
  satellites' contribution. (D): Similar as in (C), but for a barrier
  thickness of 60 layers in the case of $E_{tunn}=28$~meV. Here
  the $0.1\times 0.1$ central part of the SBZ is focused on, since the
  satellites are negligible. A double-peak structure, with a dip at
  $\bar{\Gamma}$, is evident for minority-spin.\label{fig:spinpol}}
\end{center}
\end{figure*}

The calculated current polarization as a function of barrier thickness
is shown in Fig.~\ref{fig:spinpol}. Again, two cases are presented,
corresponding to (A) mid-gap tunneling ($\EF=E_v + 190$~meV within the
barrier) and (B) shallow tunneling close to the valence band edge
($\EF=E_v + 28$~meV within the barrier), as was the case in
Fig.~\ref{fig:conduct}. These were chosen as representative of
different physical situations, where the dominant contribution to the
current stems from different parts of the Brillouin zone. In both
cases the band alignment in the injection region (i.e., after the
barrier) was chosen such that $\delta:=\EF -E_c =28$~meV. Apart from
the total current polarization, two contributions of special interest
are shown: one at exactly the $\bar{\Gamma}$-point
($P_{\bar{\Gamma}}$) and one integrated around the
$\bar{\Gamma}$-point ($P_{cntr}$). In terms of the $\kpar$-resolved
current, $I_{\uparrow}(\kpar)$ and $I_{\downarrow}(\kpar)$, these
contributions are defined as
\begin{equation}
P_{\bar{\Gamma}}=
\frac{I_{\uparrow}(\kpar=0) - I_{\downarrow}(\kpar=0)}
{I_{\uparrow}(\kpar=0) + I_{\downarrow}(\kpar=0)}
\label{eq:2}
\end{equation}
and
\begin{equation}
P_{cntr}= \frac{\int\,d^2k_{\parallel}\left[I_{\uparrow}(\kpar) -
I_{\downarrow}(\kpar)\right]} {\int\,d^2k_{\parallel}
\left[I_{\uparrow}(\kpar) + I_{\downarrow}(\kpar)\right]}
\label{eq:3}
\end{equation}
where the latter integration takes place in the central part of the
Brillouin zone where the current is non-zero (the central hot spot of
Fig.~\ref{fig:Si_pockets}).

We first discuss Fig.~\ref{fig:spinpol}A. Here the polarization is
dominated by the satellite contributions for all thicknesses, as the
current at the SBZ center is negligible
(cf.~Fig.~\ref{fig:conduct}A). The polarization is rather insensitive
to the barrier thickness, being around $P=60\%$. In
Fig.~\ref{fig:spinpol}C, the $\kpar$-resolved conductance is shown for
both spin directions in the full SBZ for a barrier thickness of
$\Ns=40$ monolayers. Evidently the SBZ center has a negligible
contribution, while the conductance at the satellite positions is
higher for majority spin than for minority spin by approximately a
factor of two. However, the contribution at exactly the
$\bar{\Gamma}$-point shows an interesting behavior, almost reaching
the ideal $P_{\bar{\Gamma}}=100\%$; the integrated value around
$\bar{\Gamma}$, $P=P_{cntr}$, is lower. Before analyzing this we
discuss Fig.~\ref{fig:spinpol}B. Here, for small barrier thicknesses,
the current is dominated by the satellites, which yield a polarization
of about 50\%. But for larger thicknesses, only the central
contribution to the current is of significance
(cf.~Fig.~\ref{fig:conduct}B). Already at a thickness of 50
monolayers, $P=P_{cntr}$. And again we see that, exactly at the SBZ
center, $P_{\bar{\Gamma}}\approx 100\%$.

In order to elucidate the situation at the SBZ center we show in
Fig.~\ref{fig:spinpol}D the $\kpar$-resolved conductance focused in
the region around the $\bar{\Gamma}$ point, for the case of shallow
tunneling ($\EF=E_v+28$~meV in the barrier) and a barrier thickness of
60 monolayers. The majority-spin conductance shows a single-peak
structure with a maximum at the $\bar{\Gamma}$ point. On the other
hand, the minority-spin conductance as a function of $\kpar$ has a
double-peak structure around $\kpar=0$, with a pronounced dip at
exactly $\kpar=0$ which is responsible for the value
$P_{\bar{\Gamma}}\approx 100\%$.

Obviously the $\bar{\Gamma}$-point enjoys special properties. This
``$\bar{\Gamma}$-point rule'' has been observed and explained in
previous works\cite{Wunnicke02,Mavropoulos02,Zwierzycki03,Wunnicke04}
on spin injection from Fe into direct gap semiconductors in the
zincblende structure (GaAs, ZnSe, InAs). The reason is traced back to
the symmetry of the wavefunctions at \EF, at $\kpar=0$. The
semiconductor wavefunctions at $\kpar=0$ have $\Delta_1$ symmetry,
which for Fe is present among the majority spin wavefunctions but
absent among the minority spin wavefunctions at \EF, at least in the
[001] direction. This symmetry mismatch of the Fe minority spin to the
semiconductor wavefunctions results in almost total reflection, so
that the current is almost 100\% polarized. Departing from $\kpar=0$,
the Si bands acquire a mixed character, such that the Fe $\Delta_5$
states (of $d_{xz}$ and $d_{yz}$ character), coupling to the Si
$\Delta_5$ complex band in the barrier, can also tunnel into the
conduction band after the barrier. Then the minority-spin transmission
rises, as is shown in Fig.~\ref{fig:spinpol}D.\cite{footnote} Note
that no such special point appears in the satellite hot spots although
their centers lie on the high-symmetry directions
$\bar{\Gamma}-\bar{M}$ (along the cubic $x$ and $y$ axes).

This behavior close to $\bar{\Gamma}$ is typical also for smaller
thicknesses and for different tunneling energies, therefore the
polarization $P_{\bar{\Gamma}}$ reaches high values also in
Fig.~\ref{fig:spinpol}A. We infer that the integrated spin polarization
can be increased if two requirements are fulfilled: (i) the central
hot spot must be as small as possible; and (ii) the satellite
contributions must be made negligible. Both can, in principle, be
fulfilled, as we now discuss.

\begin{figure}
\begin{center}
\includegraphics[width=8cm]{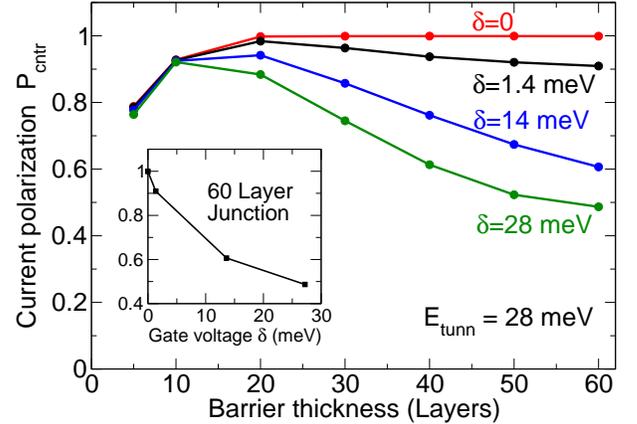}
\caption{(color online) Polarization around the $\bar{\Gamma}$ point,
  $P_{cntr}$ [see Eq.~(\ref{eq:3})], for small values of the injection
  energy $\delta$. Inset: Zero-gate ``anomaly'' of the polarization
  for a 60-layer barrier. \label{fig:zerobias}}
\end{center}
\end{figure}

Concerning requirement (i), the radius $k_{max}$ of the hot spots
depends on the injection energy $\delta=\EF-E_c$ in the injection
region as $\delta\sim k_{max}^2$. But $\delta$ is adjustable, e.g. by
tuning the gate voltage of the doping concentration. In particular for
Si, due to the high effective mass the value $\delta=28$~meV used in
the calculations hitherto is rather high and was chosen in order to
reveal the structure of the $\kpar$-resolved conductance, as already
commented in Sec.~\ref{sec:method}. By choosing a smaller $\delta$,
the SBZ center is approached more and more, and the integrated
polarization $P_{cntr}$ rises. This is demonstrated in
Fig.~\ref{fig:zerobias}. At $\delta=1.4$~meV, $P_{cntr}$ is already
over 90\%, while in the limit $\delta=0$ we obtain
$P_{cntr}\rightarrow P_{\bar{\Gamma}}\approx 100\%$. Interestingly
this results in a ``zero-gate anomaly'' (if $\delta$ is considered to
be a gate voltage), demonstrated in the inset of
Fig.~\ref{fig:zerobias} where $P_{cntr}$ is shown as a function of
$\delta$ at a barrier thickness of 60 layers. Evidently the
polarization drops abruptly with increasing $\delta$. Note that
controlled injection at about $\delta\approx 1$~meV requires low
temperatures to avoid thermal broadening, since 1~meV corresponds to
11.6~K. In this respect, the $\bar{\Gamma}$-point rule is easier
satisfied in direct-gap semiconductors, where due to the low effective
mass the $\kpar$-resolved current is confined to a tiny region around
$\bar{\Gamma}$ also at higher $\delta$.

Although Fig.~\ref{fig:zerobias} shows results for shallow tunneling
at $E_{tunn}=28$~meV, the behavior of $P_{cntr}$ is entirely
analogous at higher $E_{tunn}$, e.g., in the case of mid-gap
tunneling. Qualitatively, what changes is only the relative importance
of the satellite contributions to the central contribution.

This brings us to the discussion of requirement (ii). Is it possible
to cut off the satellites? We already saw that this happens as the
Fermi level approaches the valence band in the barrier region, if the
barrier is thick enough. But there is also another possibility, namely
by tetragonally straining Si. In tetragonally strained Si the
degeneracy of the six conduction pockets is lifted.  In the case that
$c/a<1$, the two pockets along the $c$ axis (i.e., $z$ axis along the
epitaxial direction in the presented geometry) are lowered in energy
compared to the four pockets along the $x$ and $y$ axes.\cite{Buca07}
Even a moderate strain of 0.5\% ($c/a=0.995$) was calculated to
lift the degeneracy by approximately 30~meV. As long as the injection
energy $\delta$ is kept under this limit, only the central conductance
hot spot will be populated in the injection region, while the
satellites will be cut off.

\section{Summary and concluding remarks. Outlook \label{sec:summary}}

Electrical injection from Fe into Si(001) through a Schottky barrier
has been shown to be theoretically possible. A detailed discussion of
the complex band structure of Si in the [001] direction has revealed
qualitative differences in the tunneling process compared to
direct-gap semiconductors. As a result of the complex band structure,
the current and polarization contributions at different conduction
pockets have been shown to vary very strongly depending on the
Schottky barrier thickness and the position of the Fermi level in the
barrier. Depending on these parameters, the injection efficiency has
been found to range between 50\% and 100\%.

In the calculations a particular junction setup was assumed, including
a number of approximations or idealizations. One approximation lies in
the shape of the Schottky barrier. It was assumed that the transition
from the Schottky region to the injection region is abrupt. However,
but for the case of very precise interface engineering (e.g.~with an
appropriate doping profile), the transition to the injection region is
more gradual. This would result in position-dependent decay
parameters, with the central part of the SBZ providing better
tunneling close to the interface (where the middle of the gap should
be at \EF) and the satellite positions being more efficient close to
the injection region (where $E_c$ is lowered towards \EF). Thus,
overall, either the SBZ center or the satellites would dominate the
tunneling current, depending on the exact shape of the barrier.

An idealization was that of an atomically flat Fe/Si interface, with
the in-plane lattice structure unaltered. While possible in principle,
in practice it can prove hard to achieve. If the two-dimensional
periodicity is violated at the interface, then the most severe
consequence (as regards the results presented in the present work) is
expected to be the absence of excellent spin filtering at the
$\bar{\Gamma}$-point. As was mentioned earlier, the extreme current
polarization stems from the symmetry mismatch of the Si and
minority-spin Fe wavefunctions at $\kpar=0$. Such symmetry arguments
do not hold any more in the absence of perfect interface epitaxy. It
has been shown,\cite{Zwierzycki03} e.g., in Fe/InAs(001) spin
injection (where the same principle holds), that increasing interface
disorder leads to a decrease of current polarization. However, a few
perfectly epitaxial Fe layers should be enough for a symmetry-induced
polarization. This has been found in an analogous case of
symmetry-induced polarization in Fe/MgO/Fe(001) tunnel
junctions.\cite{Heiliger07} Furthermore, it should be noted that,
since an MgO barrier is known to be selective of the $\Delta_1$ states
at $\kpar=0$, it can also be used to increase the efficiency in spin
injection experiments. Efficient spin injection has been found in
FeCo/MgO/GaAs and FePt/MgO/GaAs junctions,\cite{Jiang05} while work in
this direction has been reported also for FeCoB/MgO/Si
junctions.\cite{Uhrmann08}

A reduction of efficiency can also be caused by the formation of iron
silicide at the interface, which can sometimes lead to non-collinear
magnetic ordering. This can be avoided by inserting a nonmagnetic
metal between Fe and Si (as was done, e.g., in
Refs.~\onlinecite{Huang07a,Huang07b}); calculations on such junctions
will be the object of future work.

\section*{Acknowledgements}
It is a pleasure to thank Prof. G. Kioseoglou and Prof. I. Appelbaum
for enlightening discussions on the current state of the art in spin
injection experiments in Si, and Prof. S. Bl\"ugel for his support and
help throughout this project.


\begin{thebibliography}{99}
\bibitem{footnote} As was discussed previous papers on Fe/GaAs and
Fe/ZnSe,\cite{Wunnicke02,Mavropoulos02,Wunnicke04} the polarization at
$\kpar=0$ may be less than 100\%, because the point group rotational
symmetry around the [001] axis is $C_{4v}$ for Fe but $C_{2v}$ for the
zincblende or diamond structure. Thus, minority-spin Fe states of
$\Delta_{2'}$ symmetry ($d_{xy}$ character) at \EF\ can weakly couple
to semiconductor states of $\Delta_1$ symmetry. This weak coupling was
found\cite{Wunnicke04} to result in resonant interface states
penetrating the Schottky barrier, which under certain conditions
reduced $P_{\bar{\Gamma}}$ by 10-20\%. Such a strong contribution from
interface states is not observed in Fe/Si(001).
\end{thebibliography}
\end{document}